 \pdfoutput=1
\documentclass[Afour,sageh,times]{sagej}

\usepackage{moreverb,url}

\usepackage[colorlinks,bookmarksopen,bookmarksnumbered,citecolor=black,urlcolor=red]{hyperref}

\newcommand\BibTeX{{\rmfamily B\kern-.05em \textsc{i\kern-.025em b}\kern-.08em
T\kern-.1667em\lower.7ex\hbox{E}\kern-.125emX}}

\usepackage{balance}       
\usepackage{graphics}      
\usepackage[T1]{fontenc}   
\usepackage{txfonts}
\usepackage{mathptmx}
\usepackage{tabularx}
\usepackage{color}
\usepackage{colortbl}
\definecolor{lightgray}{gray}{0.9}
\usepackage{array}
\newcolumntype{L}[1]{>{\raggedright\let\newline\\\arraybackslash\hspace{0pt}}m{#1}}
\newcolumntype{C}[1]{>{\centering\let\newline\\\arraybackslash\hspace{0pt}}m{#1}}
\newcolumntype{R}[1]{>{\raggedleft\let\newline\\\arraybackslash\hspace{0pt}}m{#1}}
\newcolumntype{B}{@{\extracolsep{0.5cm}}c@{\extracolsep{0pt}}}%

\usepackage{booktabs}
\usepackage{textcomp}

\usepackage{multirow}

\usepackage{microtype}        
\usepackage{ccicons}          
\usepackage{lscape}
\usepackage{graphicx}

\usepackage[normalem]{ulem}

\newcommand\clearrow{\global\let\rowmac\relax}
\clearrow

\definecolor{linkColor}{RGB}{6,125,233}

\def\volumeyear{2023}

\begin{document}

\runninghead{Jhaver et al.}

\title{Decentralizing Platform Power: A Design Space of Multi-level Governance in Online Social Platforms}

\author{Shagun Jhaver\affilnum{1}, Seth Frey\affilnum{2}, and Amy X. Zhang\affilnum{3}}

\affiliation{\affilnum{1}Rutgers University, USA\\
\affilnum{2}University of California, Davis, USA\\
\affilnum{3}University of Washington, USA
}

\corrauth{Shagun Jhaver, School of Communication and Information, 
Rutgers University,
New Brunswick, NJ,
USA}

\email{sj917@rutgers.edu}

\begin{abstract}
Many have criticized the centralized and unaccountable governance of prominent online social platforms, leading to renewed interest in platform governance that incorporates multiple centers of power. 
Decentralization of power can arise horizontally, through parallel communities, each with local administration, and vertically, through multiple hierarchies of overlapping jurisdiction. Drawing from literature on federalism and polycentricity in analogous offline institutions, we scrutinize the landscape of existing platforms through the lens of \textit{multi-level governance}.
Our analysis describes how online platforms incorporate varying forms and degrees of decentralized governance. 
In particular, we propose a framework that characterizes the general design space and the various ways that \textit{middle levels} of governance vary in how they can interact with a centralized governance system above and end users below. This focus provides a starting point for new lines of inquiry between platform- and community-governance scholarship. By engaging themes of decentralization, hierarchy, power, and responsibility, while discussing concrete examples, we connect designers and theorists of online spaces. 
\end{abstract}


\maketitle

\section{Introduction}

In the months following his purchase of Twitter, Elon Musk triggered an exodus of millions of users to competing platforms; for instance, an estimated  2M users, about 1\% of its user base, made accounts on a federated competitor Mastodon \citep{peters2022twitter}.
These departures were partly in response to controversial changes to Twitter policies~\citep{insider_masto}.
Twitter is a centrally governed platform, 
where moderation decisions are made by those in power at the company and carried out unilaterally to Twitter's millions of users.
In contrast, several of the platforms to which Twitter users migrated (e.g., Mastodon, Bluesky) are designed to offer more decentralized governance. For instance, users on Mastodon can move between thousands of instances with different governance arrangements and moderation rules.
This \textit{middle level} of local administration formed through its federated architecture gives Mastodon a different approach to privacy, safety, growth, censorship, autonomy, management, and other fundamental governance properties~\citep{rozenshtein2022moderating}.




Increasingly, users are weighing questions of governance design when deciding what platforms to join, in addition to common criteria such as affordances for posting, the user base, or the platform's culture.
As platforms experiment with forms of governance beyond simply centralized, more precise terminology is necessary to differentiate between platforms designed to decentralize governance in different ways, particularly as the term ``decentralization'' carries many meanings.
As an example, while Mastodon may seem clearly different from Twitter in terms of having multiple centers of power, how does it compare to Reddit?
Much like Mastodon, Reddit is organized into many communities with a variety of governance arrangements, moderators, and rules.
However, Reddit still has a powerful \textit{top level}, enabling it to take platform-wide actions, such as banning users and communities and meddling in local governance, which would be impossible on Mastodon. 
These differences are consequential, as was recently seen when the CEO of Reddit, Steve Huffman, pushed top-down changes to combat a protest among Reddit community moderators~\citep{reddit_strike}.

The distinctions between different flavors of decentralized governance arrangements can be characterized using a lens of \textit{multi-level governance}, where end users, community moderators, and platforms sit at different levels.
Thus, in this research, we ask: what is the range of variation in how platforms organize multi-level governance systems, and what are the implications of this range for the value users receive from their platforms and the content they volunteer to contribute?




\begin{figure*}
\begin{minipage}{\textwidth}
 \centering
\includegraphics[width=0.8\linewidth]{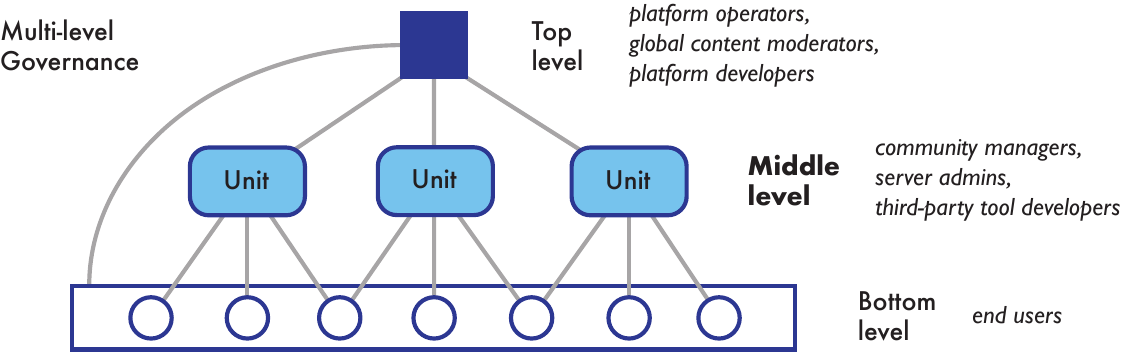}
 \captionof{figure}{A depiction of multi-level governance in an online platform, with a top, middle, and bottom level and typical actors within each level. 
 End users (denoted by the circles) are in the bottom level and are governed by one or more governance units in the middle level. All end users are additionally governed by the top level. Each unit in the middle level is also governed by the top level.
  Top levels can be concentrated and powerful (as in the cases of YouTube and Facebook) or offer a looser set of constraints (as in the cases of Wikipedia and Mastodon). Similarly, middle-level units can have a lot of governance capabilities (as in the cases of Minecraft servers and Reddit subreddits) or have only a few (as in the case of third party shared blocklists on Twitter).
 }
 \label{fig:govattr}
\end{minipage}
\end{figure*}

The increased attention to platform governance and power struggles at platforms we have seen in recent years mirrors the growing influence of online social platforms on society.
Over the last three decades, much of social media activity has consolidated around a handful of platforms that host content for billions of people~\citep{denardis2015internet}.
More recently, scholars and regulators have expressed alarm at the centralized and opaque nature of these major platforms, including their lack of procedural fairness and accountability~\citep{chen2020decentralized,fan2020digital}. 
Furthermore, some assert the futility of crafting a single set of rules that can be consistently applied over a large and diverse user population, invariably harming groups that do not fit a global standard~\citep{seering2020reconsidering,jhaver2019automated}.

In reaction, experts have called for greater decentralization of platform power to enable a plurality of moderation choices for users~\citep{fukuyama2021save,rozenshtein2022moderating}.
Despite, or perhaps because of its currency, the term decentralization is overloaded with multiple senses. For example, is a technically decentralized architecture with a clear charismatic central figure centralized? Is a strongly centralized architecture with stronger member political engagement decentralized? 
In this work, we are primarily concerned with differences in governance \textit{design}, or the ways in which governance is intended to be carried out according to platform creators, designers, and implementers. Importantly, design encompasses technical architecture as well as intentional administrative processes, which can more readily change. 
For instance, a platform may be centrally hosted but choose not to exert top-level power over local communities. In some cases, this is bound by a set of bylaws, such as with the Wikimedia Foundation; in other cases, the boundary is undefined yet tacitly agreed upon and can be renegotiated, such as with Reddit.


We summarize the major ways in which multiple centers of power are introduced to a platform's design. 
One way to achieve greater decentralization of platform power is to introduce intermediate levels of local administration~\citep{zuckerman_ch}.
Reddit offers semi-autonomous `subreddits' governed by volunteer users, while YouTube channel owners can moderate comments on their videos.
A second way is to have existing platforms enable APIs, plugins, or in-platform subscriptions to foster a marketplace of governance services~\citep{2021-modpol,frey2019emergence,fukuyama2021save}.
This marketplace allows users to choose from a suite of possible governing methods.
For instance, Twitter blocklisting tools allowed users to subscribe to the blocklist of their choice~\citep{jhaver2018blocklists,geiger2016}, serving as a weak form of local administration that only governs account blocking. More recently, Bluesky  introduced user-led moderation and curation tools, such as custom feeds and mutelists, within the platform~\citep{bluesky_mod}.
Finally, decentralized governance can arise through the use of a shared technical specification of social exchange,
such as peer-to-peer and federated protocols (e.g., Mastodon uses the ActivityPub protocol), or a shared set of data, such as blockchain technologies, including cryptocurrencies and decentralized autonomous organizations, or DAOs~\citep{wright2015decentralized}. 

In cases where platform governance incorporates decentralization, challenges still emerge.
The different governance units sometimes operate in isolation \citep{jhaver2019automated,caplan2020tiered}, resulting in redundant problem-solving and wasted effort. 
And without a central authority, local units face greater challenges addressing networked harassment like non-consensual sexual imagery~\citep{masnick_2019,marwick2018drinking}.
The specific design details of \textit{how} platforms carry out decentralization of governance are vital to achieving success 
and sustainability. 
In this work, we characterize the diverse landscape of decentralized governance in online social platforms. 
By cataloging contemporary platforms within a design space, we reveal 
the designs that remain to be explored. 

\subsection{Contributions}

We contribute a set of five dimensions to characterize the design landscape of decentralized governance on existing online social platforms.
We borrow primarily from the political science literature \citep{armitage2008governance,bache2004multi}, which has used the frame of \textit{multi-level governance} to describe decentralization in the form of authoritative decision-making dispersed vertically across multiple levels and horizontally over many local governance units~\citep{hooghe2001multi}. 
%

%
In our characterization of multi-level governance in online social platforms, we focus our attributes primarily on the \textit{middle levels} of governance.
At middle levels, many interactions play out with end users at the lowest level of a platform and with centers of power, authority, or ownership that comprise the topmost level (Figure~\ref{fig:govattr}).  
We focus on middle levels because they are the main arena for low-level agents to collectively organize with or against each other, with or against top-level agents, and even to interact with other kinds of middle levels.

As it turns out, middle levels are ubiquitous in online social platforms. We draw upon contemporary examples in this work, including Facebook Groups, Twitter shared blocklists, Reddit subreddits, YouTube and Twitch channels, Mastodon nodes, Minecraft servers, WhatsApp Groups, and Wikipedia language editions toward an understanding of the general design space of multi-level social systems online.\footnote{For simplicity, we use the term `platforms' to describe these examples as a whole, although some of these are not traditionally considered platforms.}
As part of proposing dimensions for how middle levels can vary in structure and design, we identify the other parts of platform structures that a general framework for multi-level design would address.
We describe not just the cross-level ``vertical'' dimensions that this manuscript focuses on but also the cross-unit ``lateral'' dimensions and within-unit and system-level dimensions (see Figure \ref{fig:categories}).

Given competition in an increasingly fragmented social media landscape~\citep{chatterjee2023frag}, we are entering an era where we expect a greater variety of governance structures will be attempted. 
Through our focus on multiple centers of power, we bring attention to how new designs can negotiate between the advantages of centralization versus decentralization toward more ethical, sustainable, and empowering online platforms.
Beyond theoretical insights, we offer practical design implications for online communities, drawing on lessons learned from offline and online institutions. 
We also discuss how a substantive research agenda can build upon our characterization of multi-level governance.




\section{Related Work} \label{sec:relatedWork}
Regarding the governance of online social platforms, scholars have primarily distinguished between platform governance and community governance. While \textit{platform governance} scholarship focuses on ``macro''-scale policy, compliance, and legal questions, 
\textit{community governance} research focuses on the efforts of users and volunteers to engage in local community norm-setting~\citep{gillespie2018custodians,grimmelmann2015virtues,seering2020reconsidering}. 
Although scholars in both communities acknowledge the importance of the other, the scholarship lacks a general framework for how they substitute, complement, or interact with each other. 
Our research offers early steps toward bridging the gap between platform and community governance scholarship by examining critical aspects of their interaction.

Though community-reliant platforms are decentralized, many incorporate a top level of governance comprised of platform operators, developers, or a global content moderation team overseeing the different communities. In addition, some platforms, such as Facebook, combine unitary and multi-level governance, with the Facebook News Feed governed via a centralized model and Facebook Groups employing a community-reliant model.
 The addition of a middle level allows end users to act as oftentimes volunteer community moderators. In this role, end-users can influence, monitor, and engage with platform operators~\citep{chen2020decentralized}. 
Decentralization also enables platform operators to leverage local information and innovations to improve the informational efficiency of their governance \citep{chen2020decentralized,ostrom1990governing}. Thus, with middle levels, platforms are more likely to attend to the welfare of all users.


\begin{table*}[]
\footnotesize
\setlength{\tabcolsep}{2pt}
\resizebox{\textwidth}{!}{%
\begin{tabular}{p{0.15\textwidth}p{0.15\textwidth} p{0.7\textwidth}}
 \toprule
\textbf{Platform} &\textbf{Middle Level} & \textbf{Description} \\ \midrule
Facebook & Groups &
  User-created communities that are moderated by volunteer members. Activity within all groups is also moderated by Facebook's global content moderation carrying out Facebook's site-wide community guidelines.\\ 
Reddit & Subreddits &
  User-created and managed communities, each containing its own community guidelines and automated moderation configurations. Subreddits may be banned or quarantined by the centralized Reddit administration. \\ 
Twitch & Channels &
  Channel owners live stream their videos, and each channel has its own set of moderators and settings for automated moderation of comments. The Twitch platform also enforces site-wide community guidelines. \\ 
  YouTube & Channels &
  Channel owners post videos and control the comments posted on their videos. YouTube's centralized moderation may take down videos that violate YouTube's community guidelines. It also has a site-wide filter that automatically moderates inappropriate comments. \\ 
Wikipedia & Language \newline editions &
  Each language edition has its own processes for governing article content in that language. These editions are supported by committees, developers, and operators within the Wikimedia Foundation. \\ 
WhatsApp & Groups &
  Each WhatsApp group has admins who moderate that group. WhatsApp platform operators can make platform-wide decisions, such as limits on forwarding, but are limited in their power to moderate due to end-to-end encryption. \\ 
World of \newline Warcraft & Guilds & Guilds are created by players, and can be banned by the platform, but are left to govern their own affairs. They vary greatly in their governance, membership, goals, and membership requirements.
   \\ 
World of \newline Warcraft & Communities & Communities are created by players, and can be banned by the platform. They replace the user governance of Guilds with automated mechanisms for unacquainted players to form pick-up groups and approximate closer-knit guilds. Communities are now the dominant middle level of WoW.
   \\ 
Twitter & Shared blocklists &
  Third-party moderation tools relying on the Twitter API to create lists of block-worthy users. The lists enabled subscribers to block accounts on the list automatically. Shared blocklists were either manually curated by a small staff of volunteers or automatically curated using algorithms. \\ 
  Bluesky & Custom feeds &
  Users can browse a marketplace of feed algorithms that are created by other users and then add custom feeds to their home view. Custom feeds typically require third-party infrastructure to ingest, analyze, and rank content in real-time, and they can be updated by the feed maintainer(s) at any time. \\ 
Mastodon & Nodes &
  Each self-hosted Mastodon node has server administrators who decide moderation policies and what other nodes to federate with. User accounts are tied to a specific node. While Mastodon as an open-source project has developers, each node can decide the code running on its server so long as it follows the shared ActivityPub protocol. \\ 
BitTorrent & Filesharing \newline communities & Self-hosted, emergent communities for filesharing, usually running forum software. Communities' requirements for membership tend to focus around management of free-riding. Communities have complete independence and can even use different filesharing clients, except that they all must use the BitTorrent protocol.
   \\ 
Minecraft & Servers & Although game owner Microsoft controls validation of user accounts, Minecraft worlds tend to be privately hosted. Administrators select plugins, world content, and have control over the membership and goals of their server, though the code itself is closed source.  \\ 
\bottomrule
\end{tabular}%
}
\caption{Examples from online social platforms of middle levels within a multi-level governance structure. The first 8 incorporate decentralization through the platform's design while being centrally hosted. Next, two examples (Twitter shared blocklists and Bluesky custom feeds) incorporates third-party tools to enable decentralization, while the last three examples involve the use of a shared technical standard connecting different servers.}
\label{table:middle-levels-description}
\end{table*}

While the literature on platform and online community governance spans multiple disciplines \citep{gillespie2018custodians,grimmelmann2015virtues}, the interplay between different power centers and jurisdiction levels has yet to be systematically characterized.
However, many individual case studies have highlighted a growing interest in interrogating this interplay. 
For instance, research on Reddit and Twitch content moderators has illuminated how they often govern multiple online communities, with tools and resources sometimes shared across communities~\citep{jhaver2019automated,matias2016civic}.
%
Others have examined how YouTubers, who govern channels at YouTube's middle level, collectively organized to pressure the platform to change its demonetization policy  \citep{tait2016youtubeisoverparty}.
\citet{frey2019emergence} show how player-run Minecraft communities self-organized around an emergent volunteer ecosystem of shared governance plugins.
In other work, researchers have shown how top levels of governance regulate units in the middle level that they believe are governing poorly~\citep{Chandrasekharan2017Stay,chandrasekharan2020quarantined}.
%
These cases suggest a broader pattern arising across different platforms regarding interactions between levels, one that may be informed by general scholarship on 
multi-level institutions from political science, which we next describe.

\section{Drawing from Offline Governance Literature}

Given the lack of online governance literature that systematically characterizes inter-level interactions, we turn to the offline governance literature and examine how its prevalent theories can map onto online platform governance toward developing our design space. 
We focus primarily on political science literature because its theories
 incorporate concepts such as conflict, competition, and contestation between and within levels.

%
There is no universally accepted governance theory in political science; the field has many overlapping theoretical discussions and debates \citep{ansell2016introduction,tiihonen2004governing}, but several have been developed for understanding multi-level governance systems. 
We draw from \textit{federalism}, which focuses on how nations divide power between a central government and local states, and \textit{polycentricity}, a general framework for how institutions with multiple centers of power compete and cooperate given overlapping jurisdictions. These frameworks provide an analytical structure for our study of multi-level governance and a means to challenge and strengthen our imagination beyond existing online examples~\citep{aligica2012polycentricity,scheuerman2004democractic}.

\subsection{Federalism}



One prominent theory relevant to our focus is federalism, a system of governance that divides a political territory into semi-autonomous states that share authority with one another and with a common central government~\citep{elazar2014,aroney2016types}. 
Each governing unit in this system can make laws directly affecting the citizens within its territorial purview~\citep{watts1999}.
Political scientists (traditionally, primarily Western and American) have taken the federated organization of the United States as a paradigm of multi-level governance. However, federations exist on a continuum. More peripheral systems have member states bound by loosely structured trade and defense alliances. The other extreme is ``administrative decentralization,'' where member states have little autonomy and behave more like the administrative units of an
%
%
organization~\citep{bednar2011,elazar1994}. In their idealized form, federal systems consist of neatly nested jurisdictions, with the top-level jurisdiction equal to the union of the non-overlapping jurisdictions of all lower-level systems. Within this framework, parallels to the online realm are apparent. For instance, Reddit is composed of the union of its distinct subreddits, each having some freedom to implement governance, and there is platform-level governance over all subreddits.  

\citet{aroney2016types} offers a taxonomy of federalism that is particularly well-suited to describe the complexities of multi-level platform governance. His taxonomy details features describing (1) whether the system is a federation or confederation (\textit{i.e.}, whether a unit can leave unilaterally or not), (2) whether it is formed by the aggregation of formerly independent governance units or the devolution of a formerly unitary government, (3) whether its different levels wield redundant or complementary powers over their members, and  (4) whether its lower-level governance units are symmetrical  or asymmetrical  in their powers, rights, and roles. 
The concepts of symmetric and asymmetric federations distinguish a body of equal states (as in the idealized U.S.) from a mix of states, commonwealths, districts, and territories with very different levels of autonomy and representation (as in the realized U.S.). 
In the next section, we use Aroney's taxonomy to propose some dimensions of our design space.

While several concepts transfer well to our analysis, the generality of this literature has been hampered by its focus on federalism in nations, at the expense of smaller (and more numerous) multi-level governance, a gap that \emph{polcentricity} was introduced to fill.




\subsection{Polycentric Governance}



In essence, polycentricity is an expression of self-governance capabilities that, over time, will produce a complex system of governance institutions~\citep{van2013research}. 
Although there is no clear consensus definition of polycentric governance, most scholars agree that it consists of: (1) multiple decision-making units with overlapping domains of responsibility, (2) that interact through a process of mutual adjustment in complex and ever-changing ways, and (3) generate a regularized pattern of overarching social order that captures efficiencies of scale at all levels of aggregation~\citep{mcginnis2016polycentric,aligica2012polycentricity}.
Researchers have developed many different models of polycentric governance systems to build greater clarity and specificity around the concept and highlight its posited advantages~\citep{carlisle2019polycentric,aligica2012polycentricity}.

The concept of polycentricity is
pivotal to the Ostrom school of institutional economics,  pioneered by the work of Vincent and Elinor Ostrom \citep{aligica2012polycentricity}. V. Ostrom  adopted the term `polycentricity' to describe governmental fragmentation in U.S. metropolitan areas~\citep{ostrom1961organization}.
Later, E. Ostrom's research on community-based collective management of natural resources became the best-known application of polycentricity to real-world settings~\citep{ostrom1990governing,ostrom2010beyond}.
Since then, several scholars have explored polycentric governance for sustaining natural resource systems~\citep{blomquist2005political,marshall2015polycentricity}.

Federalism can be viewed as a type of polycentricity~\citep{aligica2012polycentricity} since it incorporates many of the critical elements of the Ostroms' theory, including multiple power centers and redundant jurisdictions~\citep{van2013research}.
However, while federal systems consist of neatly nested jurisdictions under a single highest center of power, polycentric systems include crosscutting `issue-specific' jurisdictions and envisage an explicit role for autonomous private corporations, voluntary associations, and community-based organizations~\citep{mcginnis2012reflections}.
This view is well-suited to inform the analysis of multi-level online platforms containing ``nested quasi-autonomous decision-making units operating at multiple scales''~\citep{folke2005adaptive}. 
We find analogs to polycentric systems on sites like Twitch and YouTube, where different channels operate independently but account for each other through cooperation, competition, conflict, and conflict resolution~\citep{ostrom1991meaning}.

\section{A Design Space of Online Multi-level Governance} \label{sec:attributes}

\begin{figure}

\includegraphics[width=0.5\textwidth]{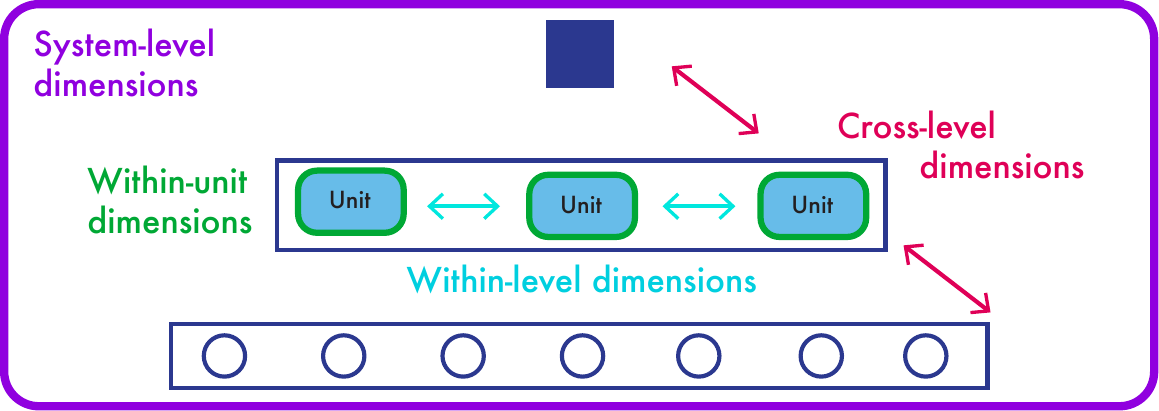}
 \captionof{figure}{Our four categories of design dimensions, including cross-level, within-level, within-unit, and system-level dimensions.
 }
 \label{fig:categories}
\end{figure}

This section presents a series of dimensions that characterize the different types of multi-level governance structures that exist within popular online social platforms. 
We curated this design space through an iterative exploratory process \citep{emmet1964learning}, combining taxonomies from federalism and polycentricity literatures with insights gleaned from case studies and our own expertise with online governance. 
This process was iterative and involved interpretation, redefinition, and verification of emerging dimensions, their comparison with one another, and their relation to prior literature.

Our analysis centers on the \textit{middle levels} of governance, including their interactions with the top and bottom levels.
Table~\ref{table:middle-levels-description} lists examples of middle levels that our design space could characterize. 
We ensured that this selection of middle levels expressed a diversity of structural features (segregated, largely centralized, third-party, etc.) and represented popular platforms focused on a wide range of topics (\textit{e.g.}, gaming, peer production) and formats (\textit{e.g.}, streaming, bulletin-board). 

We strengthened the practical utility of our emerging design space by applying it to middle levels in these thirteen implementations (Table~\ref{table:middle-levels-description}).
For this, each author familiarized themselves with each platform and its governance approach by installing its site (when necessary) and reviewing the site features and moderation documentation. 
Next, the co-authors independently considered how each platform's governance structure can be characterized through our design space. 
Following this, we compared our characterizations, and especially focused on resolving disagreements. During this process, we updated the dimension definitions to further reduce ambiguities.

%
%

\begin{figure*}
\begin{minipage}{\textwidth}
 \centering
\includegraphics[width=0.8\linewidth]{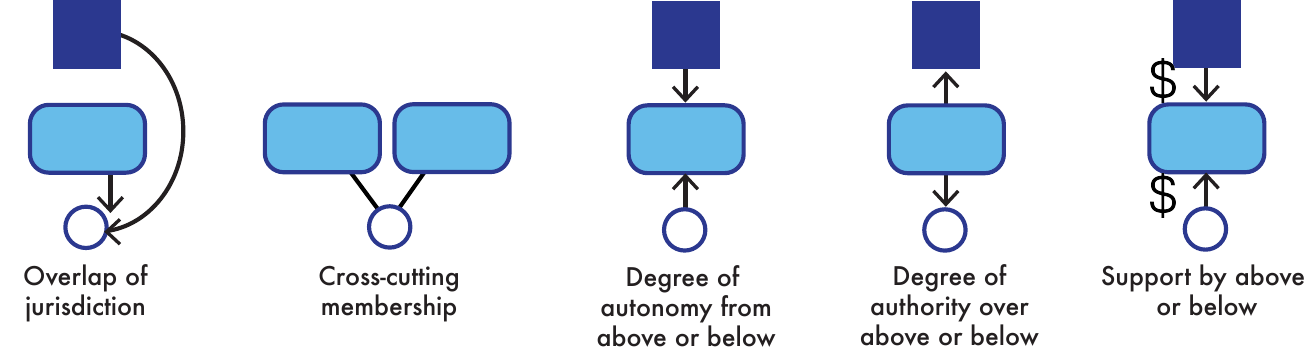}
 \captionof{figure}{Cross-level design dimensions. These dimensions characterize the relationships and interactions between middle-level units and the top level or between middle-level units and end users.
 }
 \label{fig:cross-level}
\end{minipage}
\end{figure*}

We identify four sources of design variation (Figure \ref{fig:categories}) in how online platforms structure multi-level governance:
\begin{enumerate}
\item \textit{Cross-level} or ``vertical'' dimensions of how middle-level units interact with other levels;
\item \textit{Within-level} or ``lateral'' dimensions of how middle-level units interact with one another;
\item \textit{Within-unit} internal dimensions of the middle-level units; and
\item \textit{System-level} dimensions of the whole platform. 
\end{enumerate}

In this work, we focus on cataloguing cross-level dimensions in detail. However, by situating them in a broader design space, we give scholars an appreciation of the range of implications and tensions that any design decision has for other parts of the system.
Next, we conceptualize the four sources of design variation before we focus our discussion on the cross-level dimensions.

By \textit{cross-level}, we mean variation in how middle-level units can interact with their platform administration above or member individuals below. We introduce and elaborate upon the following five cross-level dimensions in the next section: (1) overlap of jurisdiction, (2) cross-cutting membership, (3) degree of autonomy from above or below, (4) degree of authority over above or below, and (5) degree of support by above or below.

By \textit{within-level}, we mean variation in middle-level units or how they relate to or interact with one another. 
For instance, Slack workspaces have software support to connect via shared channels, as do Wikipedia articles for linking across language editions, while Mastodon nodes can federate with or block other nodes.
Variation also arises from ``transit costs'' between units: one can easily subscribe and unsubscribe to subreddits, while in WoW ``lateral exit'' is more costly~\citep{kurrild2010exit}.
Multi-level architectures can also differ in whether their middle-level units are of the same ``type'': while subreddits are mostly symmetric in the formal powers available to them, YouTube provisions its Channels differently depending on whether they are run by amateurs, legacy media organizations, or contracted producers \citep{caplan2020tiered}.

By \textit{within-unit}, we mean cross-platform differences in the internal characteristics of middle-level units with implications for their multi-level architecture. 
One example is whether the middle-level units are part of the formal architecture of the platform or emergent from users. Twitch channels, Mastodon nodes, and most other middle levels we have encountered belong to the former type. Examples of the latter type include Twitter shared blocklists and BitTorrent communities, which were created through collective community action with the help of API hacks and third-party services.
Another example of unit variation across platforms is the transparency of a middle-level unit’s governance: Reddit communities have publicly viewable rules and rule enforcement actions by default, while these governance features are private by default in Facebook Groups. 

By \textit{system-level}, we mean high-level design dimensions that influence multi-level architecture. One example is the depth of nesting of middle levels in each platform: subreddits are the only formal level between Reddit and its users, whereas Wikipedia users have organized middle levels below the level of the language edition, such as WikiProjects. Another example of system-level variation is the support for multiple orthogonal types of middle levels: WoW has two forms of middle levels, the Guild and the Community.

\subsection{Cross-level Design Dimensions} \label{sec:cross-level}
We now conduct a deep dive into a key part of the general framework, the cross-level dimensions (see Figure \ref{fig:cross-level}) that characterize the relationships and interactions (1) between middle-level units and the top-level administration and (2) between middle-level units and end users in the bottom level.

\subsubsection{Overlap of Jurisdiction.}
This dimension considers the degree of overlap between the middle and the top level regarding their areas of responsibility: what are each level's governance actions, and how much do those overlap?
Overlap of jurisdictions quantifies the redundancy in governance actions between different levels, and it is a crucial feature of polycentricity \citep{aligica2012polycentricity,carlisle2019polycentric}.
\citet{ostrom1973can} described governance systems with overlapping jurisdictions as ``highly federalized'' political systems. 
As an example, YouTube content creators can remove comments on their videos, just as YouTube platform operators can also remove comments on any video on YouTube.
Here, the responsibilities and actions afforded to the middle level are a large subset of what is afforded to the top level, leading to a high overlap in the style of a nested federation.

In the cases of low overlap, we typically observe a `thinner' top level that has access to or chooses to exert few actions compared to the middle level. The middle level does not share the responsibilities the top level oversees. 
For instance, while Mastodon node administrators are in charge of almost every aspect of governance in each node, the top level mainly is involved in code development. 

\subsubsection{Cross-cutting Membership.} 
This dimension indicates whether the membership in each unit is exclusive---that is, whether two units can govern the same user. In WhatsApp, groups have inclusive membership---users can simultaneously join more than one group. In contrast, the WoW guilds are middle-level units with exclusive membership---characters can be part of only one guild at a time. 
Cross-cutting memberships distinguish the neat hierarchical nesting of federalism from the more unconstrained notion of polycentricity, which permits multiple middle-level units to assert jurisdiction over the same user. 

When a platform has cross-cutting membership, a user's activity typically still is required to have a location in one unit or another that then determines the applicable local government. However, users may be banned from a unit due to their activity in another unit that is externally visible, so they may still experience some compounding constraints from multiple memberships. An example of this is the sometimes cascading effect of being banned from one Mastodon node or of being added to one oft-replicated Twitter shared blocklist, as units may look to other units' governance actions for guidance~\citep{jhaver2018blocklists}. Another example is bulk bans on Reddit, where moderators in one subreddit have at times banned all users who have posted on another subreddit to guard against brigades~\citep{datta2019extracting}.

\subsubsection{Degree of Autonomy from Above or Below.} 
This dimension measures the extent to which middle-level units can operate independently or have accountability to another level that constrains their autonomy. 
It maps directly to a central concern of polycentricity and federalism scholars: the degree of freedom from control by other jurisdiction levels \citep{loughlin2013routledge}.
Many platforms have middle levels that possess moderately high autonomy. 
For example, moderators of subreddits independently make a vast majority of subreddit-related decisions; they are not closely monitored by Reddit administrators.


As \citet{marshall2015polycentricity} notes,  \textit{de facto} autonomy may matter as much as formal autonomy. 
While platforms may promise greater autonomy to middle-level units, they may still employ governing-at-a-distance strategies to exercise greater control over outcomes, for example, by imposing reporting and compliance requirements \citep{marshall2015polycentricity,carlisle2019polycentric}.
Another strategy is through the threat of sanctions.
Middle levels have less autonomy when they are held to account by platform operators who can ban middle-level units that do not adhere to platform policies.\footnote{\url{https://www.facebook.com/policies/pages_groups_events/}} 
In some cases, technical affordances restrict this assertion of power.
For example, in the case of WhatsApp,  end-to-end encryption limits platform visibility into the actions of middle-level administrators.

We also conceptualize the degree of autonomy of a middle level from those below it. In the `implicit feudalism' model inherent on many social platforms~\citep{schneider2021admins}, a community's founder has absolute autonomy over that community by default and is not directly responsible to users.  Often, the only accountability is indirect, through the threat that users will exit for other more accountable or otherwise desirable communities~\citep{frey2020effective}. The more rare and interesting case is a middle-level formally accountable to those below it.
On Wikipedia, the community has the power to withdraw an administrator's special privileges in cases of abuse of authority, restrict their use of certain functions, and place them on probation.\footnote{\url{https://en.wikipedia.org/wiki/Wikipedia:Administrators}}
Another case is when users can report abusive community moderators to the platform, at which point the platform uses its power to provide accountability; this is currently supported on Reddit.


\subsubsection{Degree of Authority over Levels Above or Below.}  A related but distinguishable dimension to a middle-level unit's autonomy \textit{from} is authority \textit{over}, which refers to the ability to regulate or sanction.
The question of authority is as central to federalism and polycentricity as autonomy \citep{morrison2019black}. Indeed, a middle-level unit having autonomy implies it has authority over something. 
However, piecing out the direction of its authority, like its autonomy, helps to distinguish different instances of multi-level governance online. 

For example, admins of Facebook Groups have authority over Group members because they have the power to sanction members or their posts. These admins also make governance decisions autonomously, as they are usually not elected by Group members. 
On the other hand, admins of a Wikipedia language edition have authority over the governance of Wikipedia pages in that language. However, they do not operate autonomously since community members can strip away their administrator rights.

Middle levels usually have significant authority over low levels, with some work demonstrating the emergence of administrator oligarchies~\citep{shaw2014laboratories}. One exception we have found is Twitter shared blocklists \citep{geiger2016}: as a third-party service with limited functionality, these blocklists  could not strongly sanction their subscribers.

Middle levels can also have formal authority over higher levels, though most examples involve informal input. 
The closest example may be the Wikimedia Foundation, which, likely due to its mission, has informal mechanisms for granting authority to its language editions. 
However, the most evident and concrete mechanism---the Foundation's partially member-elected board---encodes a direct democracy that empowers bottom-level users rather than a representative democracy that might empower middle-level units. As another example, the Reddit Mod Council\footnote{\url{https://mods.reddithelp.com/hc/en-us/articles/4415446939917-Reddit-Mod-Council}} enables a select group of subreddit moderators to informally advise platform operators; however, there is no mechanism for sanctioning the top level.



\subsubsection{Support by Levels Above or Below.}
On some platforms, middle levels are supported by higher levels through various means, such as receiving technical help or getting access to moderation resources. For example, Twitch provides streamers on each channel moderator tools that let them set rules to remove inappropriate content automatically. Discord launched a Moderator Academy for educating server mods.\footnote{\url{https://discord.com/moderation}}
Middle levels can also be supported by lower levels, such as the private servers of Mastodon, Minecraft, or WoW, which sometimes depend on users' financial support. 

Relevant indicators of levels of support from above could include whether a middle level is self-hosted or hosted on a central platform and whether it receives funding, human staff support, or special access to developer tools from the top level.
Additionally, levels of support from below could be quantified based on whether a middle level is wholly or partially user-run and user-funded and has tools or culture supporting voluntary contributions.

\section{Design Implications}


As can be seen from our characterization of the landscape, multi-level governance can take many forms online. 
Decentralization has the potential to create governance more attuned to community needs. 
However, it can also mean more points of failure and more significant overhead and inefficiency~\citep{bednar2008robust}. 
We point to design implications of our analysis of multi-level governance for platforms and communities. 

\subsection{Enabling Innovation, Adaptation, and Healthy Competition}
One benefit of decentralization is the ability of governance units to innovate, as emphasized in the phrase ``laboratories of democracy'' used to describe state governments in the U.S.
As communities evolve, their survival partly depends on their willingness to evolve their governance in response to events. 
The polycentricity literature emphasizes continual adaptation in changing environments~\citep{ostrom1999coping,carlisle2019polycentric}, which depends on centers having the capability to continually experiment with rules \citep{carlisle2019polycentric,ostrom1999coping}. 
Thus, to enable innovation and adaptation, online social platforms should provide autonomy and otherwise support local administrators' experiments with community guidelines, sanctioning criteria, or automation settings \citep{jhaver2019automated,matias2018civilservant}.
Competition for users is another outgrowth of increased decentralization that can lead to better user conditions. New communities often form when a faction of an existing community is dissatisfied with its moderation ~\citep{matias2016civic,mcgillicuddy2016}. 
According to polycentricity and federalism scholars, competition between different units induces self-regulating tendencies as it compels units to demonstrate their value to members~\citep{ostrom1961organization,thiel2019governing,mcginnis2012reflections}. 
However, competition can be detrimental for users when it leads to cross-community conflict or significant consolidation. Platforms with cross-cutting memberships may be able to avoid
excessive competition, 
as joining one unit does not imply exiting another. This permits units to be complementary despite the appearance of competitive overlap.

\subsection{Facilitating Social Learning}
Many governance scholars identify social learning and building social capital as essential conditions for increasing resilience and sustainability ~\citep{berkes2010devolution,folke2005adaptive,gelcich2014towards}, without which each center must arrive at an optimum arrangement through trial and error \citep{ostrom1999coping}. 
However, many platforms do not offer formal avenues for administrators in different centers to communicate with one another, leading to inefficiencies such as the building of redundant tools~\citep{jhaver2019automated}. 
Allowing for cross-cutting membership permits members to be in multiple communities and enables knowledge transfer.
In their analysis of Wikia, \citet{zhu2014impact} found that when users participate in multiple communities, their membership helps the survival of each due to the transfer of best practices. 

Separately, the top level can also facilitate social learning. Given some overlap in the jurisdiction, the top level could promote specific middle-level solutions to additional middle-level units. For example, after many subreddits voluntarily used Reddit Automod, an automated moderation tool, Reddit realized its potential and incorporated it as a default mechanism on all subreddits~\citep{jhaver2019automated}. 
Another example is the Discord Moderator Academy, which included lessons, an exam, and an online community of fellow Discord moderators before being shut down in 2022.\footnote{\url{https://discord.com/blog/announcing-the-discord-moderator-academy-exam}}

\subsection{Establishing Accountability}

Holding decision-makers accountable for poor performance is crucial for the proper functioning of governance systems \citep{carlisle2019polycentric}.
Surprisingly, we find little evidence of top levels actively evaluating the moderation practices of middle-level units, except in the most public or egregious cases. 
One rare example is Twitch's proposed ``Brand Safety Score,''\footnote{\url{https://www.digitalinformationworld.com/2021/03/twitch-scores-users-through-brand.html}} which attempts to score streamers based on aspects including how they govern their stream. However, it seems primarily for the benefit of advertisers and not accountability to members.
Accountability can also come from the levels below. 
Processes for users to report, elect, or otherwise exert pressure on local administrators would exert higher accountability to lower levels~\citep{frey2020effective}.
A middle level supported from below can work autonomously from the level above and would be obliged to hold greater accountability to the level below for its decisions.

\section{Limitations and Future Work}
These dimensions are far from the only insights that scholarship on offline multi-level institutions can offer to online platforms. We aim to start a conversation that more closely links political science and public administration scholars to the platform governance community. Our dimensions are not measurable indicators but qualitative descriptions sufficiently general enough to be contextualized to each platform. We call for further studies to precisely operationalize these dimensions and carefully develop the broader design space's within-level, within-unit, and system-level dimensions. 
While we focus on middle levels in this work, examining units in other levels can also reveal valuable insights. For example, \citet{jhaver2023personalizing} show how end-users deploy platform-offered AI tools to personalize the governance they experience.

We hope that a substantive research agenda can build upon the design dimensions we have presented here.
Some research questions that represent important next steps include: 
How can platforms support governance units in their experiments with a diversity of governance strategies? 
Which design mechanisms and policies can help platforms foster healthy competition between different governance units? 
What initiatives, technical means, and tools can platforms develop to facilitate social learning? 
How can platform levels with overlapping jurisdiction collaborate effectively to ensure that no governance unit free rides on others’ efforts or oversteps clearly defined bounds? 
How do different incentives and sanctions levied on governance units affect their operations in the long run?

\section{Conclusion}
Social systems are often complex, interdependent, and have a hierarchical structure~\citep{simon1991architecture}. Understanding their complexity requires studying multi-level design, which has been explored in multiple disciplines. Multi-level frameworks are particularly important for examining large-scale social platforms, as evident in the ubiquity and diversity of middle levels that we find. 
In this work, we wrangle the many manifestations of multi-level online platforms into a generative design space. 
Drawing upon multi-level offline governance theories, we examine our framework's design implications for more empowering and accountable governance.
Unlike offline institutions, online platforms are an ideal laboratory for exploring the design space of potential multi-level governance architectures as designers can iterate quickly, base decisions on mass data, and compare communities directly. 

\bibliographystyle{SageH}
\bibliography{refs.bib,moderation_references.bib}
\end{document}